\documentclass[article,prl,twocolumn,showpacs]{revtex4}
\usepackage{graphicx,amsmath,amssymb,color}
\usepackage[utf8]{inputenc}
\usepackage{mathrsfs}
\usepackage{amsfonts}
\usepackage{caption}
\usepackage{enumerate}
\usepackage[toc,page]{appendix}
\usepackage{url}
\usepackage[normalem]{ulem}
\usepackage[ocgcolorlinks,colorlinks=true,linkcolor=blue,citecolor=red]{hyperref}
\usepackage{latexsym}
\usepackage{amsthm}
\usepackage{verbatim}
\usepackage{psfrag}
\usepackage{fancyhdr}
\usepackage{pdflscape}
\usepackage{lipsum}

\DeclareMathOperator{\Tr}{Tr}

\newcommand{\ket}[1]{\left| #1 \right\rangle}                                   

\begin{document}
	\title{Spatial versus Sequential Correlations for Random Access Coding}
	\author{Armin Tavakoli$^{1,2}$, Breno Marques$^{1,3}$, Marcin Paw{\l}owski$^2$, Mohamed Bourennane$^1$}
	
	\affiliation{$^{1}$Department of Physics, Stockholm University, S-10691 Stockholm, Sweden.\\ 
		$^2$Institute of Theoretical Physics and Astrophysics, Uniwersytet Gda\'nski, PL-80-952 Gda\'nsk, Poland.\\
		$^3$Instituto de Física, Universidade de São Paulo, P.O. Box 6 6318, 05315-970 São Paulo, Brazil}

	
	\date{\today}
	
	
	\begin{abstract}
		Random access codes are important for a wide range of applications in quantum information. However, their implementation with quantum theory can be made in two very different ways: (i) by distributing data with strong spatial correlations violating a Bell inequality, or (ii) using quantum communication channels to create stronger-than-classical sequential correlations between state preparation and measurement outcome. Here, we study this duality of the quantum realization. We present a family of Bell inequalities tailored to the task at hand and study their quantum violations. Remarkably, we show that the use of spatial and sequential quantum correlations imposes different limitations on the performance of quantum random access codes: sequential correlations can outperform spatial correlations. We discuss the physics behind the observed discrepancy between spatial and sequential quantum correlations.

	\end{abstract}
	
	
	
	\maketitle

	
	\textit{Introduction.---}
	Quantum theory can break the limitations of classical physics on the strength of correlations arising in space-like separated measurement events (spatial correlations). Spatial correlations without a classical explaination are recognized as such if they can be shown to violate a Bell inequality \cite{B64}. Such non-classicality tests can be directly re-formulated as games \cite{BZPZ04} in which the involved parties that perform the measurements are given some random inputs by a referee and asked to return outputs satisfying particular constraints. If they succeed, they win the game.

	A class of games that has recieved significant attention is a communication task known as a random access code (RAC). A RAC is a collaborative task in which a party Alice holds a data string $\bar{x}\!=\!x_0...x_{n-1}\!\!\in\!\!\{0,...,d-1\}^n$, while another party Bob decides on a value $y\!\in\!\{0,...,n-1\}$ indicating that he wants to access the particular element $x_y$ in $\bar{x}$. The RAC consists of the partnership employing a strategy in which Alice communicates at most $\log d$ bits of information from which Bob can find $x_y$ with a high average probability $p^C$. We denote such RACs by $n^{(d)}\!\!\rightarrow \!1$.

	Quantum solutions to RACs can increase the success probability beyond the classical bound. Quantum RACs are important for a broad range of applications  including finite automata \cite{ANTV}, network coding \cite{HI}, quantum state tomography \cite{A07}, device independent dimension witnessing \cite{WCD08, AB14}, semi-device independent cryptography \cite{PB11} and randomness expansion \cite{HW11}, nonlocal games \cite{MT14}, tests of contextuality \cite{SB09}, studies of no-signaling resources \cite{GH} and characterization of quantum correlations from information theory \cite{IC09}.

	To perform a quantum RAC one must distribute data with suitable correlations. For instance, by sharing an entangled state, Alice and Bob could generate spatial correlations that violate a suitably choosen Bell inequality in such a way that when supplied with $\log d$ bits of communication, Bob can compute $x_y$ with an average probability $p^E\!\!>\!p^C$. The simplest example of such an entanglement-assisted random access code (EARAC) is the CHSH Bell-type inequality \cite{CHSH69}, for which the corresponding game \cite{W10} is as follows. Alice and Bob are given inputs $x,y\!\in\!\{0,1\}$ respectively and asked to return outputs $a,b\!\in\!\{0,1\}$ such that $a+b-xy\!=0\!\!\mod{2}$. In the standard interpretation of the CHSH game there is no communication between the players; the referee checks the success condition using $(a,b)$. In the RAC interpretation, Bob will check the success condition and to this purpose Alice will therefore communicate her binary outcome $a$ to him. In order to win, Bob needs Alice to send him the bit $z_0\equiv a$ whenever $y\!=\!0$, and the bit $z_1\!\equiv a-x\!\mod{2}$ whenever $y\!=\!1$. If Bob indeed obtains $z_y$, the game is won since  $z_y+b=0\!\mod{2}$. It is well-known that using a shared classical random variable, the partnership can do no better than $p^C\!=\!0.75$ whereas they can reach $p^E\!\approx\! 0.854$ by performing measurements on a shared entangled state \cite{BCD01}. Therefore, the performance in the CHSH game directly corresponds to the success probability of a $2^{(2)}\!\!\rightarrow \!1$ RAC.

	However, strong spatial correlations are not the only way in which quantum theory can improve RACs. The $n^{(d)}\!\!\rightarrow \!1$ RAC can also be performed by replacing the classical communication channel with a quantum channel of the same capacity i.e. Alice encodes her data into a quantum $d$-level system which is sent to Bob who performs a measurement determined by $y$ from which he obtains $x_y$ with average probability $p^Q$. Such quantum communication random access codes (QCRACs) can for instance be used to reproduce the success probability of the EARAC in the CHSH game. The power of the QCRAC stems from strong sequential correlations, arising in 'prepare-and-measure' schemes, stored in the resulting probability distribution of Bob obtaining a particular outcome for given measurement and state preparation.

	At first sight, it may appear as if the spatial and sequential correlations giving rise to EARACs and QCRACs are two sides of the same coin. This is however false. In fact, it is known that for any game with binary answers and $d\!=\!2$, spatial quantum correlations perform at least as well as sequential quantum correlations \cite{PW12}, and there are explicit examples of such EARACs outperforming QCRACs \cite{PZ10}. This raises the question of understanding the dual interpretation of quantum RACs.

	Here, we will construct both spatial correlation inequalities (Bell inequalities) and sequential correlation inequalities tailored to RACs and study their respective violations by quantum theory, associated to EARACs and QCRACs. In particular, we study violations dependig on dimension of Hilbert space. Interestingly, we show that when $d>2$, sequential quantum correlations can outperform spatial quantum correlations in RACs, thus reversing the comparative strength of the EARAC and QCRAC known for $d=2$ \cite{PW12, PZ10}.
	
\begin{figure}
		\centering
		\includegraphics[width=0.82\columnwidth]{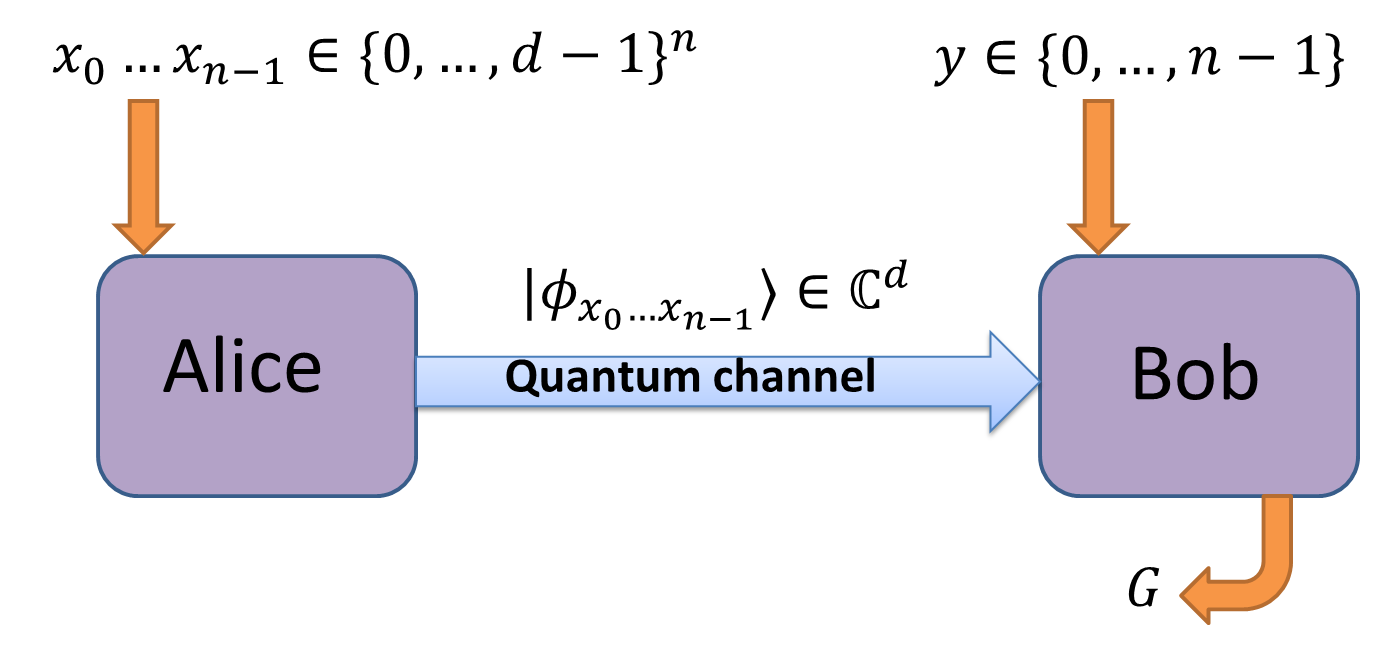}
		\caption{The $n^{(d)}\rightarrow 1$ QCRAC. Alice determines one of $d^n$ states which she sends to Bob who performs a measurement $y$ and outputs the outcome as his guess $G$.}
		\label{figseq}
	\end{figure}

	
	\textit{Sequential correlation inequalities for RACs.---} We shorly summarize the work on $n^{(d)}\rightarrow 1$ RACs using sequential correlations in Ref.\cite{THMB15}. From a given data set $x_0...x_{n-1}$, Alice prepares a quantum state $|\phi_{x_0...x_{n-1}}\rangle\in \mathbb{C}^d$ which is communicated to Bob over a quantum channel. Bob makes one of $n$ measurements $y$ on $|\phi_{x_0...x_{n-1}}\rangle$ and uses the output $G$ as his guess for $x_y$, see figure \ref{figseq}.
	The sequential correlation inequality for the RAC can then be written
	\begin{equation}\label{pQ}
	p^Q_{n,d}=\frac{1}{nd^n}\!\!\!\sum_{x_0,...,x_{n-1}=0}^{d-1}\sum_{y=0}^{n-1}P(G=x_y|x_0,...,x_{n-1},y)\leq p^C_{n,d}.
	\end{equation} 
	where the bound $p^C_{n,d}$ is the optimal success probability \cite{AKR15} achievable for classical $n^{(d)}\!\!\rightarrow \!1$ RACs, as given explicitly in Ref.\cite{THMB15}. 
	
	We shortly outline the quantum protocol for $n=2$. Introduce the operators  $X=\sum_{k=0}^{d-1}|k\rangle\langle k-1|$ and $Z=\sum_{k=0}^{d-1}\omega^{k}|k\rangle\langle k|$ with $\omega=e^{2\pi i/d}$. The eigenstates of $Z$ ($X$) are the elements of the computational basis $\{|l\rangle\}_{l=0}^{d-1}$ (the Fourier transformed elements of the computational basis: $|e_l\rangle=\frac{1}{\sqrt{d}}\sum_{k=0}^{d-1}\omega^{kl}|k\rangle$). Alice's $d^2$ state preparations $|\phi_{x_0x_1}\rangle$ for $x_0,x_1=0,...,d-1$ are chosen as
	\begin{equation}
	|\phi_{00}\rangle=\frac{|0\rangle+|e_0\rangle}{\sqrt{2+\frac{2}{\sqrt{d}}}} \hspace{10 mm}	|\phi_{x_0x_1}\rangle =X^{x_0}Z^{x_1}|\phi_{00}\rangle .
	\end{equation}
	If Bob is interested in $x_0$ ($x_1$) he performs a mesurement in the basis $\{|l\rangle \}$ ($\{|e_l\rangle\}$) obtaining an outcome $G\in\{0,...,d-1\}$. Evaluating \eqref{pQ} it is straightforward to find the average success probability $p^Q_{2,d}=1/2+1/2\sqrt{d}$. These specific QCRACs were shown to be optimal (at least) for $d=2,3,4,5$ \cite{THMB15}. In comparison, the optimal classical success probability is strictly smaller: $p^C_{2,d}=1/2+1/2d$ \cite{THMB15}.

	
	\begin{figure}
		\centering
		\includegraphics[width=\columnwidth]{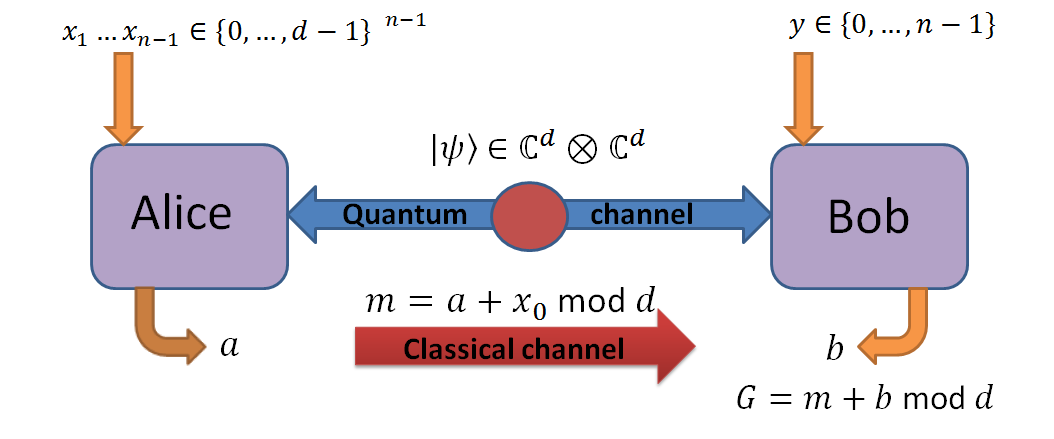}
		\caption{The $n^{(d)}\!\!\rightarrow \!1$ EARAC. Alice determines one of $d^{n-1}$ measurements and from the outcome constructs and sends message $m$ to Bob who performs a measurement $y$ with an outcome $b$  from which a guess $G$ is outputted.}
		\label{fig:1}
	\end{figure}
	\textit{Bell inequalities for RACs.---} $n^{(d)}\!\!\rightarrow \!1$ RACs can be constructed using spatial correlations. Let Alice and Bob share a state $|\psi\rangle\in \mathbb{C}^d\otimes \mathbb{C}^d$ and let Alice perform a measurement $x$ chosen among $d^{n-1}$ options indexed by $n\!-\!1$ elements of her data set; $x\equiv x_1...x_{n-1}$. The measurement returns an outcome denoted $a\in\{0,...,d-1\}$. Alice computes the message $m\equiv a+x_0\!\mod{d}$ and communicates it to Bob. Bob performs one of $n$ measurements determined from his input $y$, which returns an outcome $b\in\{0,...,d-1\}$, which he uses to output  $G\equiv m+b\mod{d}$ as his guess for the value of $x_y$, see figure \ref{fig:1}. Therefore, Bob succeeds in accessing $x_y$ whenever the condition $a+b=x_y-x_0\mod{d}$ is satisfied. Since we are interested in the average success probability of performing the RAC, we must normalize the probability of satisfying the success condition over the number of possible measurement settings. The success probability, $p^{E}_{n,d}$, of the $n^{(d)}\!\!\rightarrow \!1$ EARAC is therefore the left-hand side of the following Bell inequality:
	\begin{equation}\label{Ineq}
	p_{n,d}^{E}\!\!=\!\! \frac{1}{nd^{n-1}}\!\!\sum_{x=0}^{d^{n-1}\!\!-1}\hspace{0.8 mm}\sum_{y=0}^{n-1}P(a+b=x_y-x_0|x,y)\leq p^C_{n,d},
	\end{equation}
	where the argument of $P(\cdot)$ is computed modulo $d$. For the particular case of $n=2$ the inequality can be simplified by Alice mapping her inputs $x_0,x_1$ into $z_0\equiv x_0$ and $z_1\equiv x_1-x_0\mod{d}$, and using the pair $(z_0,z_1)$ as the input data of Alice in the RAC. We can then  replace the condition $a+b=x_y-x_0\mod{d}$ in \eqref{Ineq} with the condition $a+b=xy\mod{d}$.  Notice that for $(n,d)=(2,2)$ this will return the CHSH inequality, as discussed in the introduction.

	A remark on the above: One should note that the $\log d$ bits of communication allows for a more general construction of the above Bell inequalities. In general, we could imagine that Alice processes her measurement outcome and her input $x_0$ by some function $f_d:\{0,...,d-1\}^{2}\rightarrow \{0,...,d-1\}$ into the message $m=f_d(a,x_0)$ which is sent to Bob. Bob then applies some function $h_d:\{0,...,d-1\}^{2}\rightarrow \{0,...,d-1\}$ and outputs $G=h_d(m,b)$ as his guess for $x_y$. If $f_d$ or $h_d$ are not linear in their arguments this will correspond to a different Bell inequality than \eqref{Ineq}. However, this would no longer correspond to realizing the random access code as an XOR game which is e.g. the form of the CHSH game, i.e. the $2^{(2)}\rightarrow 1$ EARAC. Furthermore, a loss of linearity leads to products between inputs and outputs in the argument of $P$. Therefore, $G$ will in general not be a surjective function in every variable i.e. as we let e.g. $a$ run over its $d$ possible values, the values of $G$ will not correspond to the full set $\{0,...,d-1\}$. However, $x_y$ is sampled from a uniform distribution and thus can attain all values in $\{0,...,d-1\}$. Thus, there will be values of $x_y$ which are impossible to attain no matter how strongly correlated $a$ and $b$ are. Due to these reasons, we will restrict to considering linear functions $f_d$ and $h_d$ as stated in \eqref{Ineq}.
	
	We emphasize that in this paper we are interested in comparing the strengths of 
		spatial and sequential correlations. Therefore, we use an object introduced in \cite{GH}, called RAC-box. It is a non-classical 
		probability distribution which, when augmented with a $\log d$ bits of 
		classical communication, can be used to create an EARAC according to the 
		recipe from Figure \ref{fig:1}. The efficiency of the EARAC is then proportional to 
		the violation of a particular Bell inequality obtained from these correlations. To this violation we will compare the QCRAC.

	
	\textit{Comparing EARACs and QCRACs.---} Since it is increasingly complicated to find analytical maximal violations of the inequalities, we have used numerical methods to investigate inequality \eqref{Ineq} for some values of $(n,d)$. We have used semidefinite programs (SDPs) \cite{VB96} in a see-saw method to optimize $p^{E}_{n,d}$ over the set of quantum measurements for Alice and Bob respectively, which returns a lower bound on $p^{E}_{n,d}$. We have also used the intermediate level, $Q_{1+ab}$, of the hierarchy of quantum correlations \cite{NPA07} to find an upper bound on $p^{E}_{n,d}$. The results are presented in Table \ref{tab:1} together with the optimal performance of the corresponding QCRAC and classical RAC.

	We notice that our method only provides an exact value, up to numerical precision, of the optimal $p^E$ when $(n,d)=(2,2),(2,3)$. This is due to the fact that there is no guarantee that the upper (lower) bound is tight. However, for every pair $(n,d)$ studied in table \ref{tab:1} the results are sufficient to prove that when $d>2$, there are QCRACs that outperform optimal EARAC. This result is the opposite to that found in Refs.\cite{PZ10, PW12}: that for $d=2$ EARACs can outperform QCRACs but never vice versa. Clearly, the dimension of Hilbert space crucially influences the comparative strenght of the two types of quantum correlations.

		\begin{table}[t]
			\centering 
			\begin{tabular}{|c|c|c|c|c|} 
				\hline 
				$(n,d)$ & $p^{C}$ & $p^{Q}$ &   $p_{num}^{E} $ & $p_{Q_{1+ab}}^{E}$   \\ [0.5ex] 
				\hline
				(2,2) & 0.7500 & 0.8536 & 0.8536 & 0.8536  \\
				(2,3) & 0.6667 & 0.7889 & 0.7778 & 0.7778   \\
				(2,4) & 0.6250 & 0.7500  & 0.7405  & 0.7441  \\
				(2,5) & 0.6000 & 0.7236 & 0.7178 & 0.7179  \\
				(3,3) & 0.6296 & 0.6971 & 0.6854 &  0.6912  \\
				\hline 
			\end{tabular}
			\caption{Numerics for EARACs compared with classical RACs and QCRACs.} 
			\label{tab:1}
		\end{table}
	
	\textit{Approximating the $2^{(3)}\rightarrow 1$ EARAC.---} We shall now explicitly derive a violation the Bell inequality which up to the numerical precision used in the simulations displayed in Table \ref{tab:1} corresponds to the performance of the optimal $2^{(3)}\rightarrow 1$ EARAC.
	
	Alice and Bob will share the maximally entangled state $
	|\psi_{max}\rangle\!=\!\frac{1}{\sqrt{3}}\left(|00\rangle+|11\rangle+|22\rangle\right)$ associated to the operator $\rho\!=\!|\psi\rangle\langle \psi|$. Bob chooses between two measurements and we label his measurement operators $B^b_y$ with $y\!=\!0$ associated to a measurement in the basis $\{|l\rangle\}$ and $y\!=\!1$ associated to a measurement in the (earlier introduced) basis $\{|e_l\rangle\}$. Additionally, we shift the labels of the outcomes of Bob when $y\!=\!1$ by setting $B_1^b=|e_{b+2}\rangle\langle e_{b+2}|$.

	Alice's measurement is determined by $x\equiv x_1$ and we denote her measurement operators $A_x^a$ for $a,x=0,1,2$. Let us choose the measurement operator $A_0^0$ as 
	\begin{equation}\label{3operator}
	A_0^0=\begin{pmatrix}
	\frac{7}{9} & -\frac{1-3i\sqrt{3}}{18} & -\frac{2+i\sqrt{3}}{9}\\
	-\frac{1+3i\sqrt{3}}{18} & \frac{1}{9} & \frac{\omega}{9}\\
	\frac{-2+i\sqrt{3}}{9} &  \frac{\omega^2}{9} & \frac{1}{9}
	\end{pmatrix}.
	\end{equation}
	Now, we use the operators $X=\sum_{k=0}^{2}|k\rangle\langle k+1|$ and $Z=\sum_{k=0}^{2}\omega^{k}|k\rangle\langle k|$ to define the remaining measurement operators of Alice by $A_x^a=\left(X^{a}Z^{a-x}\right)A_0^0\left(X^{a}Z^{a-x}\right)^\dagger$. The operators $A_x^a$ can be written $A_x^a=\sum_{k,k'}\omega^{(a-x)(k-k')}\lambda_{k,k'}|k-a\rangle\langle k'-a|$
	where we have denoted the elements of $A_0^0$ by $\lambda_{k,k'}$. It is straightforward to verify that $\{A_x^a\}_a$ for $x=0,1,2$ is a valid quantum measurement.
	
	We compute the probability distribution $P(a,b|x,y)$, beginning with all entries with $y=0$: 
	\begin{multline}\label{2y0}
	P(a,b|x,y=0) =\!\Tr\left(\rho A_x^a \otimes B_0^b\right)\! =\!\Tr\Bigg(\frac{1}{3}\!\!\sum_{l,l'=0}^{2} |l'l'\rangle\langle ll|\\
	\!\!\!\!\!\!\!\!\!\!\!\!\times\!\!\!\!\!\! \sum_{k,k'=0,1,2}\omega^{(a-x)(k-k')}\lambda_{k,k'}|k-a,b\rangle\langle k'-a,b|\Bigg)\\
	= \frac{1}{3}\sum_{k,k'}\omega^{(a-x)(k-k')}\lambda_{k,k'} \delta_{b,k-a}\delta_{b,k'-a}
	=\frac{\lambda_{a+b,a+b}}{3}.
	\end{multline}
	Secondly, we compute all entries of $P(a,b|x,y)$ with $y\!=\!1$:
	\begin{multline}
	P(a,b|x,y=1)= \!\Tr\left(\rho A_x^a \otimes B_1^b\right)\! =\!\frac{1}{9}\Tr\Bigg(\sum_{l,l'=0}^{2}|l'l'\rangle\langle ll|\\
	\times\!\!\sum_{k,k'=0}^{2}\!\!\omega^{(a-x)(k-k')}\lambda_{k,k'}|k-a\rangle\langle k'-a|
	\otimes \!\!\sum_{j,j'=0}^{2}\!\!\omega^{(b+2)(j-j')}|j\rangle\langle j'|\Bigg)\\
	=\frac{1}{9}
	\sum_{k,k'=0,1,2}\lambda_{k,k'}\omega^{(k-k')(a+b-x+2)}.
	\end{multline}
	
	Now, we use $P(a,b|x,y)$ to compute the success probability $p^E_{2,3}$ directly from  \eqref{Ineq},
	\begin{multline}
	p^{E}_{2,3}\!=\!\frac{1}{6}\!\sum_{x=0}^{2}\sum_{y=0}^{1}P(a+b\!=\!xy|x,y)\!= \!\frac{1}{6}\!\sum_{x=0}^{2}\Bigg(\!\sum_{a+b=0} \frac{\lambda_{a+b,a+b}}{3}\\
	+\!\frac{1}{9}\!\!\sum_{a+b=x}\sum_{k,k'=0}^{2}\omega^{(k-k')(a+b-x+2)}\lambda_{k,k'} \Bigg)=
	\frac{\lambda_{0,0}}{2}\\
	+\frac{1}{6}\left(1+2\Re\left(\omega^2\lambda_{1,0}+\omega\lambda_{2,0}+\omega^2\lambda_{2,1}\right)\right)
	=\frac{7}{9}\nleq p^C_{2,3}=\frac{2}{3}. 
	\end{multline}
	The obtained success probability $p^E_{2,3}=7/9$ coincides, up to numerical precision, with the upper bound presented in Table \ref{tab:1} and therefore either very well approximates or directly corresponds to the optimal $2^{(3)}\rightarrow 1$ EARAC.

	\textit{A proposed experimental realization of the $2^{(3)}\rightarrow 1$ EARAC.---} 
	We will now present an experimental setup for implementing the $2^{(3)}\!\!\rightarrow \!1$ EARAC. Such a setup can have applications in e.g. quantum communication complexity, device independent random number generation and cryptography. 
	
	Even though the associated QCRAC yields a larger success probability, an experimental proposal of the EARAC, and its implementation in the future, is of interest due to two reasons: (i) the optimal $2^{(4)}\!\!\rightarrow \!1$ QCRAC has been realized in Ref.\cite{THMB15} and the experimental technique can be modified to realize the $2^{(3)}\!\!\rightarrow \!1$ QCRAC, and (ii) EARACs have an important advantage over QCRACs in the sense that multiple copies of a basic EARAC can be used to implement a significantly more complex EARAC.  Therefore, complex EARACs can be implemented without increasing the qualitative experimental complexity in terms of the requirements on state preparation and measurements for the basic $2^{(3)}\!\!\rightarrow \!1$ EARAC.  For example, extending the technique of Ref.\cite{PZ10}, it is not too difficult to show that using two copies of the optimal $2^{(3)}\!\!\rightarrow \!1$ EARAC, one can construct a $4^{(3)}\!\!\rightarrow \!1$ EARAC with success probability $p^{E}_{4,3}=(7/9)^2+2(1/9)^2\approx 0.6296$ which clearly outperforms the classical bound $p^C_{4,3}\approx 0.5926$. 
	
	\begin{figure}
		\hspace{0.1cm}
		\begin{center}
			\includegraphics[width=\columnwidth]{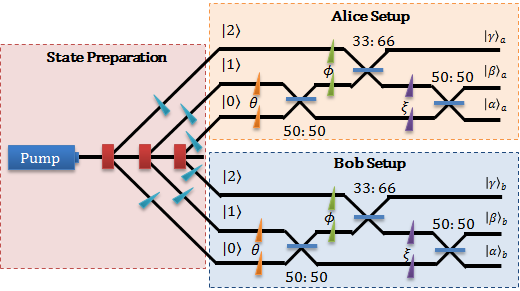}
		\end{center}
		\caption{Experimental proposal for the optimal $2^{(3)}\rightarrow 1$ EARAC. The state preparation utilizes three nonlinear crystals wheras the measurements are performed by a combination of BSs and phase plates. The photons are detected in coincidence by single-photon detectors.}
		\label{abraco}
	\end{figure}
	
	We have implemented the optimal $2^{(3)} \!\!\rightarrow\! 1$ EARAC by encoding the information in the three photon paths. The preparation of the state $|\psi_{max}\rangle$ is illustrated in the red box shown in Figure \ref{abraco}. A laser pumps three successive nonlinear crystals that generate photon pairs with equal efficiency. The state created by the crystal $\ell\in\{0,1,2\}$ is $\ket{\ell}_a\ket{\ell}_b$. Keeping all relative phases between the arms constant, the maximally entangled state is created. After the state preparation Alice (orange box) and Bob (blue box), implement their measurements as shown in Figure \ref{abraco}. Alice and Bob use similar setups with two 50:50 and one 33:66 beam splitter (BS) together with phase plates used to control phases $\theta$, $\phi$ and $\xi$ between the input paths for the first, second and third BS respectively. To perform a local measurement, Alice and Bob set their phases as given in Table \ref{cruzeiro}. The states $\ket{\alpha}$, $\ket{\beta}$ and $\ket{\gamma}$ correspond to the states projected onto by $A^0_x$, $A^1_x$ and $A^2_x$ for Alice and $B^0_y$, $B^1_y$ and $B^2_y$ for Bob. The photons are detected by single-photon detectors and the results are given by the coincidences between the Alice and Bobs paths.
	
	Observe that the implemented measurements are not the same as those presented in the theory but rotated with a unitary action.

	\begin{table}[h!]
		\centering 
		\begin{tabular}{|c||c|c|c|} 
			\hline 
			Measurement & Phase $\theta$ & Phase $\phi$ & Phase $\xi$\\ [0.5ex] 
			\hline
			$y=0$ & 0 & 0 & 0 \\
			\hline
			$y=1$ & 0 & $2\pi/3$ & 0\\
			\hline
			$x=0$ & $-5\pi/6$ & $5\pi/6$ & 0\\
			\hline
			$x=1$ & $5\pi/6$ & $-5\pi/6$ & 0\\
			\hline
			$x=2$ & $-\pi/2$ & $-\pi/2$ & 0\\
			\hline 
		\end{tabular}
		\caption{The settings of the phases corresponding to the measurements of Alice and Bob.} 
		\label{cruzeiro}
	\end{table}

	\textit{Discussion.---} Let us now discuss the physics behinds the observed discrepancy between the performance of EARACs and QCRACs. For simplicity, we consider the $2^{(3)}\rightarrow 1$ RAC which we know is, up to numerical precision, optimally implemented with spatial correlations using a maximally entangled state of local dimension $3$.

	The intuition for the advantage of the QCRAC over the EARAC is that entanglement imposes an additional constraint on the possible local states of Bob. In QCRACs Alice's nine preparations can be any quantum states with dimension $3$. An optimal EARAC must involve sharing the maximally entangled state since if Bob's outcomes can be at least partially predicted by him beforehand, they carry some information about the state and less than the optimal information about Alice's input. Therefore, an additional constraint appears: a measurement of Alice renders Bob's qutrit in one of three orthogonal states. Thus, Bob's nine possible states must be three sets of three orthogonal states. Therefore, the QCRAC must perform at least as well as the EARAC. In fact, given any one of Alice's nine preparations $|\psi_j\rangle$ in the QCRAC, the value $|\langle \psi_{j}|\psi_{j'}\rangle|$, for $j\neq j'$, takes one of only two possible non-zero values. Thus, there is significant symmetry in Alice's preparations, but no two preparations are orthogonal. Clearly it is impossible to achieve this set of local states of Bob in the optimal EARAC.

	However, why does not the above constraint of entanglement make a differece for the $2^{(2)}\rightarrow 1$ EARAC, which is also optimally implemented with local dimension $d$? The reason lies in the geometry of Hilbert space, namely the Bloch sphere. We allow Alice to arrange her four preparations freely on the Bloch sphere, and similarly for Bob's two pairs of orthogonal states (associated to his respective measurement outcomes). The optimal arrangement is to arrange Bob's four states on a Bloch sphere disk so that they form the vertices of a square, and arrange Alice's states on the same disk so that they also form a square but rotated relative to Bob's square by $\pi/2$. The eight states thus form an octagon on the Bloch sphere disk. Therefore, Alice's optimal preparations are two pairs of orthogonal states. However, in the $2^{(3)}\rightarrow 1$ EARAC, such a symmetric arrangement of Alice's and Bob's states is not possible due to the less symmetric geometry of three dimensional Hilbert space. Instead we encounter optimal arrangements of the type discussed in the previous paragraph. Indeed, it was proven in Ref.\cite{PW12} that if the communication and outcomes of Bob are binary, QCRACs would never outperform EARACs. However, it was also shown \cite{O10} that the optimal measurements for such EARACs are constructed from the generators of Clifford algebras. If we want to find optimal EARACs for $d$-level communication and outcomes the straightforward choice would be to use generalization of these algebras with higher dimensional generators. These do exist \cite{C1} but they lack the nice symmetric properties of the original ones which makes them not very useful for EARACs.

	For EARACs of $d>3$, we do not know that the local dimension of the optimal entangled state coincides with $d$. A thorough study of the trade-off between the orthogonality restriction on the possible local states of Bob and the local dimension of the entangled state would be of interest. This is however beyond the scope of our work. 
	
	Furthermore, the above may appear contradictory since we earlier stated that it is known that the EARACs can outperform QCRACs if $d=2$ and $n\geq 4$ \cite{PZ10}. However, the construction of \cite{PZ10} is not based on a single use of a Bell inequality violation, but uses concatenations of $2^{(2)}\rightarrow 1$ and $3^{(2)}\rightarrow 1$ EARACs to construct $n^{(2)}\rightarrow 1$ EARACs, for which our above argument is clearly no longer valid. Indeed, if one attempts to realize e.g.  a $4^{(2)}\rightarrow 1$ EARAC with a single two qubit entangled state, one will not be able to outperform the QCRAC as explained by our argument.

	
	\textit{Conclusions.---} We have studied the relation between spatial and sequential quantum correlations as resources for performing quantum RACs. Interestingly, we found that despite spatial correlations being a more powerful resource in two-dimensional Hilbert space \cite{PZ10}, the opposite relation can be found for larger Hilbert space dimensions. This significantly enriches the complexity of the more general problem of understanding the limitations of spatial and sequential correlations in quantum games. For a given communication game, how can one determine if spatial or sequential quantum correlations perform the best? Furthermore, stronger sequential quantum correlations are certainly of interest for various quantum information protocols. 	
	
	As a more technical open problem, we mention that due to limited computational power we have been unable to find both upper and lower bounds on EARACs and QCRACs except for small $n$. It would be interesting if EARACs were to re-gain the upper hand when $n$ is large enough, since this may be suggested by the intuition behind the maximal number of mutually unbiased bases in $\mathbb{C}^d$.
	
	Finally, although our work is constrained to bipartite systems generalizations to multipartite systems may be of interest. We will continue our research in that direction.

	\begin{acknowledgements}
		We thank Ad\'an Cabello for interesting discussions. We thank Andreas Winter for sharing his knowledge on Clifford algebras. This project was supported by the Swedish
		Research Council, ADOPT, FAPESP and CAPES (Brazil) and NCN grant no. UMO-2014/14/E/ST2/00020.  
	\end{acknowledgements}

\end{document}